%% file: article.tex
\documentclass[aps,pre,preprint,superscriptaddress]{revtex4-2}
\usepackage{graphicx}
\usepackage{import}
\usepackage{xcolor}
\usepackage{indentfirst}
\usepackage{float}
\usepackage{amsmath, amssymb}
\usepackage{siunitx}

\DeclareSIUnit\angstrom{\text{Å}}

\begin{document}

% Use the \preprint command to place your local institutional report
% number in the upper righthand corner of the title page in preprint mode.
% Multiple \preprint commands are allowed.
% Use the 'preprintnumbers' class option to override journal defaults
% to display numbers if necessary
%\preprint{}

%Title of paper
\title{Structure and Nonlinear Index of Refraction of Sunset Yellow Lyotropic Chromonic Liquid Crystal in the Isotropic and Nematic Phases}

\author{Dennys Reis}
\email[Contact author: ]{dennys.reis@usp.br}
\affiliation{Instituto de Física, Universidade de São Paulo, São Paulo, SP, Brazil}

\author{Renato Mafra Moysés}
\author{Lino Misoguti}
\affiliation{Instituto de Física de São Carlos, Universidade de São Paulo, São Carlos, SP, Brazil}

\author{Antônio Martins Figueiredo Neto}
\affiliation{Instituto de Física, Universidade de São Paulo, São Paulo, SP, Brazil}

\date{\today}

\begin{abstract}
Lyotropic chromonic liquid crystals are formed by the self-assembly of aromatic compounds in concentrated solutions. Despite numerous applications of chromonic systems in optical and photonic devices, they all make use of the anisotropic linear optical properties of the nematic or columnar liquid crystalline phases. This paper extends the investigations of chromonic systems to the domain of nonlinear optics. For this purpose, the magnitude and sign of the nonlinear refractive indices, $n_2,$ were measured by the nonlinear ellipse rotation (NER) technique. This was performed on aqueous solutions of sunset yellow azo dye, the prototypical chromonic system. Samples with different concentrations and temperatures were used, both in the isotropic and nematic phases. In addition, the molecular aggregation states of the chromonic samples as a function of temperature and concentration were investigated by wide angle X-ray scattering. NER measurements as a function of the laser pulse width from $65\,fs$ to $\sim 5\,ps$ allowed the decomposition of $n_2$ into a fast contribution, $n_{2,fast},$ associated with molecular electronic processes, and a slow one $n_{2,slow},$ associated with molecular reorientational processes. It was shown that $n_{2,fast}$ doubled from the isotropic phases of the $15$ to the $30\,\%\,\text{w/w}$ samples, proportionally to the increase in mass fraction. However, $n_{2,fast}$ for the aligned nematic phase of $30\,\%\,\text{w/w}$ sample was higher than the double of the corresponding value for the $15\,\%\,\text{w/w}$ sample, showing an effect associated to the orientational order of this phase. Also, $n_{2,fast}$ was shown to depend linearly on temperature.

% insert abstract here
\end{abstract}

% insert suggested keywords - APS authors don't need to do this
%\keywords{}

%\maketitle must follow title, authors, abstract, and keywords
\maketitle

% body of paper here - Use proper section commands
% References should be done using the \cite, \ref, and \label commands
\section{ Introduction }
% Put \label in argument of \section for cross-referencing
%\section{\label{}}
%\subsection{}
%\subsubsection{}

Lyotropic chromonic liquid crystals are liquid crystalline phases observed in solutions of some aromatic compounds at appropriate temperature, pressure and relative concentrations \cite{dierking_lyotropic_2024}. These mesophases have nanostructures, mechanism of formation and properties distinct from other lyotropic mesophases, such as the amphiphilic lyotropic mesophases. Therefore, they are called chromonics \cite{lydon_chromonic_2010, lydon_chromonic_2011, collings_nature_2015, zhou_lyotropic_2017}. 

The molecules that form chromonic mesophases have a central core composed by aromatic rings, which constitute a planar molecular structure, and also have polar functional groups at their edges, which make them soluble. Typically, these functional groups are ionic or hydrophilic, given that most reported chromonic systems are aqueous solutions. Chromonic mesophases have been observed in solutions of drugs, dyes, nucleic acids, among others molecules \cite{tam-chang_chromonic_2008}.

In solution, these molecules spontaneously self-assemble into anisometric columnar stacks by noncovalent attractive interactions between the aromatic cores of adjacent molecules  \cite{lydon_chromonic_2010, chami_molecular_2010}. Under thermodynamic equilibrium conditions,the stacks exhibit an aggregation number distribution that depends on temperature and relative concentrations of the compounds, among other factors \cite{collings_nature_2015}. The average length of columns increases with the concentration of aromatic compound \cite{joshi_concentration_2009}. The molecular stackings are the basic structural units of the chromonic mesophases.

Also, there are evidences that the self-assembly process is almost isodesmic \cite{dickinson_aggregate_2009, zhang_extensive_2022}. The stacking free energy is $ \approx 8 \, k_B T,$ according to UV-Vis spectroscopy measurements in dilute aqueous solutions of disodium cromoglycate (DSCG) and of the azo dye sunset yellow \cite{zhang_extensive_2022}.

Typically, in addition to the isotropic liquid phase, two liquid crystalline phases were identified in chromonic systems \cite{lydon_chromonic_2010}. In order of increasing aromatic compound concentration in solution, the sequence of phases observed is: isotropic (I), nematic (N), and hexagonal (M). 

In the nematic phase, the molecular stacks exhibit long-range orientational order, but no long-range positional order. I.e., the stacks tend to be positioned as in the liquid state, yet oriented parallel to a specific direction, represented by a unit vector $\hat{n}$ called the director. The $\hat{n}$ and $-\hat{n}$ states of the director are indistinguishable.

The structures of the stackings in each chromonic mesophase were investigated previously by X-ray scattering \cite{horowitz_aggregation_2005, joshi_concentration_2009, xiao_structural_2014, eun_lyotropic_2020}, UV-Vis absorption spectroscopy \cite{dickinson_aggregate_2009, zhang_extensive_2022}, nuclear magnetic resonance \cite{edwards_chromonic_2008, renshaw_nmr_2010}, and computational simulations \cite{chami_molecular_2010, potter_self-assembly_2020}. 

Over the last two decades, chromonic systems have been used in various optical and photonic devices. These include the construction of polarizers operating across the entire visible spectral \cite{tam-chang_chromonic_2008}; polarizers made from microstructured substrates \cite{tam-chang_template-guided_2006}; real-time holographic gratings using the nematic phase of an azo dye \cite{hahn_investigation_1998}; optical retardation waveplates using a N* phase \cite{lavrentovich_planar_2003}; and biosensors for medical diagnostics \cite{shiyanovskii_real-time_2005}, among others.

Virtually all applications exploit the anisotropies of the linear optical properties of the N or M phases, as well as of solid films produced by drying these aligned mesophases. However, to the best of our knowledge, no applications or investigations on the nonlinear optical properties of the chromonic mesophases were found in the literature.

The aim of this paper is to extend the investigations of the structural and nonlinear optical properties of chromonic systems. This was carried out with samples of aqueous solutions of the dye sunset yellow (SSY), the prototypical and most studied chromonic systems. The nanostructures were investigated for samples of different concentrations, temperatures and phases using wide angle X-ray scattering (WAXS). For the nonlinear optics investigation, the magnitude and sign of the nonlinear refractive indices, $n_2,$ of chromonic samples were measured by the nonlinear ellipse rotation (NER) technique \cite{boyd_chapter_2008, maker_intensity-dependent_1964, maker_study_1965, liu_nonlinear_2007, miguez_accurate_2014, miguez_nonlinear_2015, miguez_measurement_2017, de_souza_measurement_2017, fernandes_small_2021, cunha_nonlinear_2025}.

\section{ Materials and Methods }

\subsection{Samples}

The aqueous solutions of the azo dye sunset yellow were prepared with different percentage by mass of SSY, $w_{SSY}.$ The samples are identified by their $w_{SSY}$ values, e.g., 30SSY refers to the sample with composition $w_{SSY} = 30\,\%\,\text{w/w SSY}$ and $w_{H_2O} = 70\,\%\,\text{w/w H}_2\text{O}.$

The substances were added and weighed in clean and dry test tubes. First, the purified dye were added, followed by the water. The mixtures were homogenized by repeated and alternating application of a vortex mixer and centrifugation. The test tubes were properly sealed to prevent evaporation.

The dye SSY was bought from Sigma-Aldrich with a purity grade of 90\%. Deionized water were obtained from a Millipore Milli-Q purification system. For reproducibility and compatibility with published results, the SSY was purified by recrystallization in three identical cycles. In each one, SSY was diluted in water to obtain a homogeneous solution close to saturation. Then, analytical grade ethanol was carefully added to the solution to obtain SSY precipitates. After a resting period, the solution was vacuum filtered, and the precipitate was dried at $50\,^{\circ}C$ for 24 hours in a vacuum oven. After three cycles, SSY with a high degree of purity was obtained. The procedure was verified by comparing the obtained phase diagram with those previously published \cite{eun_lyotropic_2020}.

\subsection{Experimental Techniques}

\subsubsection{ Polarized Light Optical Microscopy}

The Leitz Ortoplan-Pol microscope was equipped with an achromatic objective of 10$\times$/0.25 NA, and with a digital camera model Guppy F-503C, from AVT. The textures of the liquid crystalline phases were observed between crossed polarizers.

The samples were introduced by capillary action into rectangular glass capillary tubes with a thickness of $200\,\mu m,$ from Vitrocom. Subsequently, the ends of the tubes were sealed with photopolymerizable resin. The tubes were then placed in an Instec HS1 sample holder for temperature control with precision of $\pm 0.05\,^{\circ}C.$

\subsubsection{ 2D Small- and Wide-Angle X-ray Scattering }

The two-dimensional small- and wide-angle X-ray scattering measurements (SWAXS) were performed using a laboratory Xenocs Xeuss 2.0 instrument. In Xeuss, the X-ray beam is generated by a Genix3D system, consisting of a \textit{microfocus} X-ray source with a copper anode and a FOX3D X-ray mirror system, the latter used for beam monochromatization and collimation. Subsequently, the beam is further collimated by two \textit{scatterless} slits. The X-ray beam incident on the sample had parallel collimation, a square cross-section with $0.7\,mm$ sides, and a wavelength of $\lambda = 1.5418 \, \mathring {\mathrm A} \, (Cu \, K_{\alpha}).$ The measurements were performed in transmission geometry. The X-ray scattering images were measured by a Dectris Pilatus 300K detector at a sample-to-detector distance of $14.3\,cm.$

The samples were introduced into cylindrical borosilicate capillaries with $1.5\, mm$ diameter, model Mark-tubes from Hilgenberg. These capillaries were previously cut at both ends and glued in a homemade metal structure with lids. Thereby, the capillaries could be cleaned, reused multiple times, and put in vacuum. Consequently, the samples, the water and the empty capillary backgrounds could be measured all in the same capillary. This ensured an accurate background subtraction in the data treatment. In addition, these reusable capillaries were put in a sample holder with temperature control with precision of $\pm 0.5\,^{\circ}C.$

For X-ray scattering patterns with circular symmetry, full azimuthal averages were performed, and the images were reduced to curves of scattered intensity, $I,$ as a function of the modulus of the scattering vector, $q = \frac{4\pi}{\lambda} \sin \theta,$ where $2\theta$ is the scattering angle. This is the case for isotropic phases.

For anisotropic liquid crystalline phases, the patterns were anisotropic and partially aligned due to the capillary filling procedure. Therefore, the azimuthal averages were performed over circular sectors of $10^{\circ}$ around the preferential directions of the patterns.

Azimuthal and radial averaging, corrections for solid angle and polarization, and subtraction of electronic noise and background scattering were performed using a homemade software developed with the pyFAI \cite{ashiotis_fast_2015} python library.

\subsubsection{ Nonlinear Ellipse Rotation }

The nonlinear polarization ellipse rotation phenomena was used to measure the nonlinear refractive index, $n_2,$ of the chromonic samples.

\begin{figure}[h]
	\centering
	\def\svgwidth{\linewidth}
	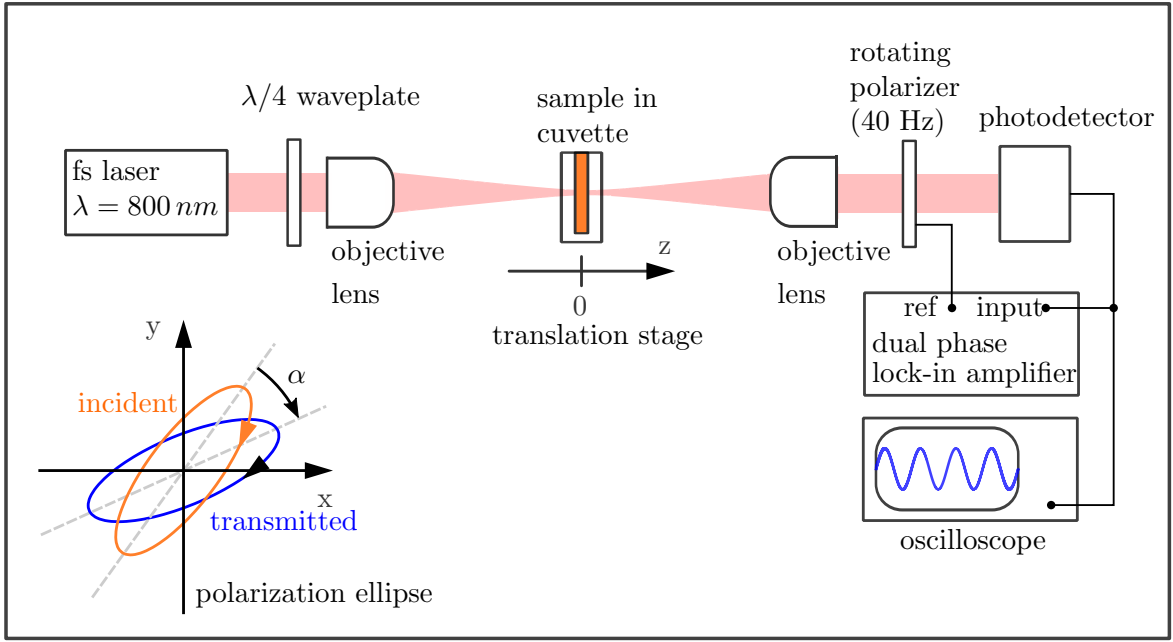
	\caption{Diagram of the nonlinear ellipse rotation (NER) setup. In the bottom left, there is an illustration of the rotation of the polarization ellipse of the transmitted beam (orange) with respect to the polarization ellipse of the incident beam (blue), due to the nonlinear rotation effect. The eccentricities of the polarization ellipses are the same in both cases. The polarization ellipse rotation angle, $< \alpha(z) >,$ is measured as a function of sample position, $z$, in beam propagation direction, around the focus position $z=0.$}
	\label{fig:ner:setup}
\end{figure}

An optical setup was used to measure the angle of rotation of the polarization ellipse, $\alpha,$ induced by a high intensity laser beam. The setup of the nonlinear ellipse rotation (NER) technique is outlined in Figure \ref{fig:ner:setup}.

The laser used was a Ti:sapphire multipass chirped-pulse-amplifier system, model Dragon from K\&M Labs. It generated light pulses of $\lambda = 800\,nm,$ at a pulse rate of $1\,kHz,$ and with tunable width from $\approx 70\,fs$ to $5\,ps.$ The pulse width was controlled by an internal pair of gratings of the pulse compressor. The bandwidth limited pulse was about $35\,fs$ (FWHM). 

The $\lambda/4$ waveplate controlled the polarization state of the incident beam. Then, the polarized beam was focused with an objective lens of $f \approx 3\,cm.$ 
The sample holder was placed on a motorized translation stage. This way, the sample in a cuvette was translated along the beam propagation direction, or z-direction, through the focus of the laser beam. The z-axis was defined such that the focus was at $z = 0,$ z negative was the pre-focal positions and z positive the postfocal positions. 

The transmitted light was collimated with another objective lens of $f \approx 3\,cm.$ and modulated with a rotating polarizer at a rate of $\approx 40 Hz.$ Finally, the modulated signal was measured with a photodetector connected to the input of a dual phase lock-in amplifier. The reference signal of the lock-in was the rotating polarizer signal measured with a frequency sensor.

From the in-phase and quadrature signals from the lock-in, the  phase of the transmitted signal was measured and, consequently, the angle of the polarization ellipse.

The polarization ellipse rotation angle as a function of sample position, $z$, in beam propagation direction, around the focus position $z=0,$ is given by \cite{miguez_accurate_2014, miguez_nonlinear_2015}:

\begin{equation}
< \alpha(z) > = \frac{1}{\sqrt{2}} \, \frac{2\pi}{\lambda} \,
           \left( \frac{ \sin{ 2 \varphi } }{ 2 } \right)
           \left( \frac{ B }{4\,n_0^2 c \epsilon_0  } \right) \, (n_0 z_0) \,I_0
           \left[ tan^{-1} \left( \frac{ z + L/(2 n_0) }{ z_0 } \right) - tan^{-1} \left( \frac{ z - L/(2 n_0) }{ z_0 } \right) \right]
\label{eq:ner:angle}
\end{equation}
where $c$ is the light speed; $\epsilon_0$ is the permittivity of free space; $\lambda$ is the wavelength of laser light; $\varphi$ is the angle between the plane of polarization of the incident laser beam and the fast axis of the $\lambda/4$ waveplate; $w_0$ is the focused beam waist and $z_0 = \frac{\pi \omega_0^2}{\lambda}$ is the Rayleigh length; $I_0$ is the peak intensity at the focal point $z=0;$ $L$ is the sample thickness; $n_0$ is the sample linear refractive index; and $B$ is a coefficient of the third-order nonlinear susceptibility tensor of the sample, related to $n_2$ \cite{boyd_chapter_2008, maker_intensity-dependent_1964, liu_nonlinear_2007}.
    
For a sample in a cuvette, the NER signal is given by a sum of terms, each one for a material medium, as shown below \cite{miguez_measurement_2017, de_souza_measurement_2017, fernandes_small_2021}:

\begin{equation} 
\label{eq:ner:sample}
<\alpha > = <\alpha>_{holder 1} + <\alpha>_{sample} + <\alpha>_{holder 2},
\end{equation}
where the ``holder'' subscripts refer to the cuvette windows.

All the samples were measured in a Quartz cuvette of $1\,mm$ path length (model QX, Hellma), except for the oriented N phase sample. This sample was measured in a liquid crystal cell with anti-parallel polyimide alignment layers of $20 \pm 2 \mu m$ path length (model LCC1324, Thorlabs). The quartz cuvette windows have known linear and nonlinear refractive index at $\lambda = 800\,nm$, $n_0 = 1.4533$ and $n_2 = 2.5 \times 10^{-20} \, m^2/W,$ respectively. Therefore, the cuvette windows were used as the reference material for the NER measurements.

\section{ Results and Discussion }

\begin{figure}[h]
\centering
\includegraphics[scale=0.17]{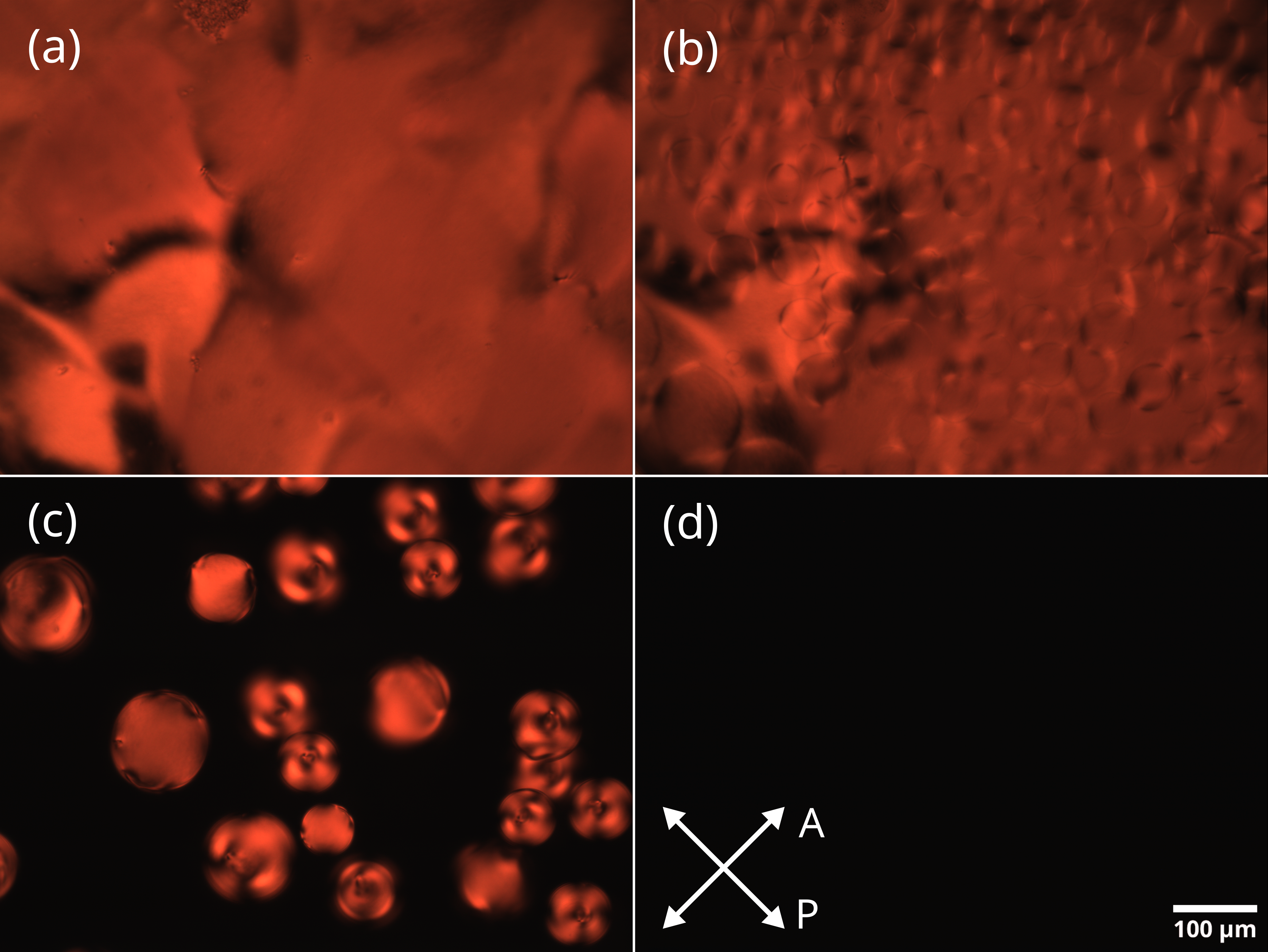}
\caption{Textures of the chromonic phases of sample 30SSY, observed between crossed polarizers by polarized light microscopy on heating. (a) N phase, at $T = 25.0\,^{\circ} C;$ (b) N and I phases coexistence, at $T = 36.0 \,^{\circ} C;$ (c) N and I phases coexistence, at $T = 45.0 \,^{\circ} C;$ and (d) I phase, at $T = 48.0\,^{\circ} C.$ P and A represent the directions of the polarizer and the analyzer, respectively.}
\label{fig:texturas}
\end{figure}

Polarized light optical microscopy was used for observation of textures and identification of phases; measurement of phase transition temperatures; construction of the phase diagram; and to check the director alignment of samples in nematic phase. Figure \ref{fig:texturas} shows the textures of the (a) nematic phase,  (b - c) nematic and isotropic phases coexistence region, and (d) isotropic phase.

\subsection{Structure}

\begin{figure}[h]
\centering
\includegraphics[scale=0.7]{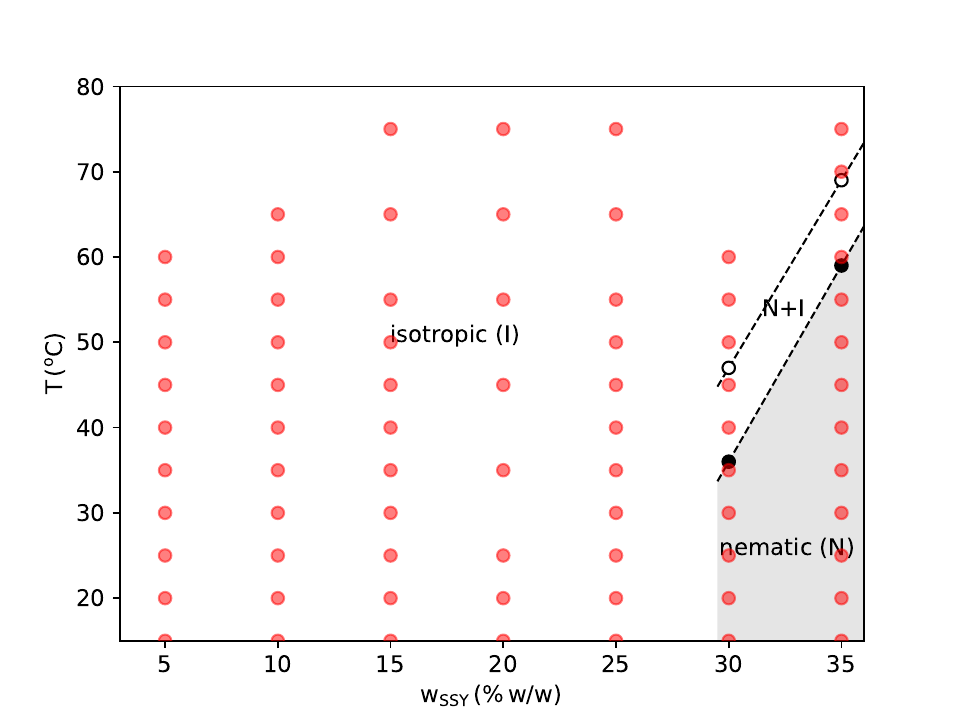}
\caption{Phase diagram for the mixture of dye sunset yellow and water, where $w_{SSY}$ is the mass fraction, in $\%\,\text{w/w SSY}.$ Samples with $w_{SSY}$ from $5$ to $35\,\%\,\text{w/w},$ in steps of $5\,\%\,\text{w/w},$ were investigated. $N+I$ represents the phase coexistence region. The dashed lines were interpolated into the data as a guide. Temperatures and concentrations of the samples measured by WAXS are marked in red.}
	\label{fig:cromonic:phase_diagram}
\end{figure}

Figure \ref{fig:cromonic:phase_diagram} shows the partial phase diagram of the SSY and water mixture. Also, the samples concentrations and temperatures investigated by WAXS are marked over the phase diagram.

\begin{figure}[ht]
	\centering
	\includegraphics[scale=0.9]{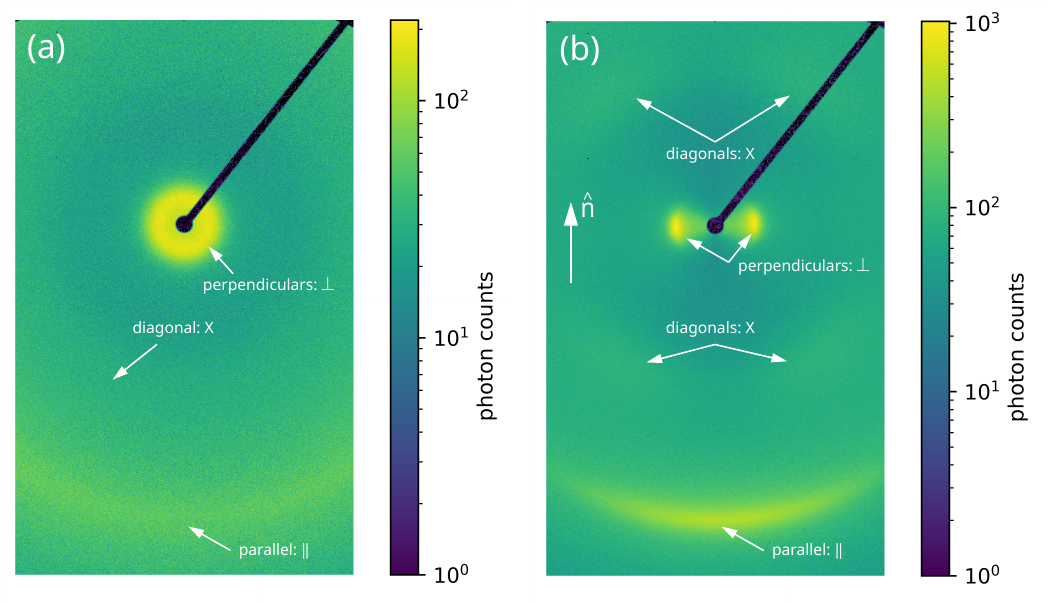}
	\caption{X-ray scattering images. (a) Isotropic phase of sample 20SSY at $T = 25.0\,^{\circ} C;$ (b) oriented nematic phase of sample 35SSY at $T = 30.0\,^{\circ} C.$ The bands are identified in each image. Perpendicular, parallel and diagonal refer to  the stacking direction, e.g., parallel refers to X-ray scattering band due to the structures along, or parallel to, the stacking direction. Corresponding bands in both images have the same name. In (b), $\hat{n}$ is the N phase director.}
	\label{fig:waxs:2D}
\end{figure}

The typical two-dimensional WAXS patterns of the isotropic and oriented nematic phases are shown in Figures \ref{fig:waxs:2D}(a) and (b), respectively. 

\begin{figure}[ht]
	\centering
	\includegraphics[scale=0.75]{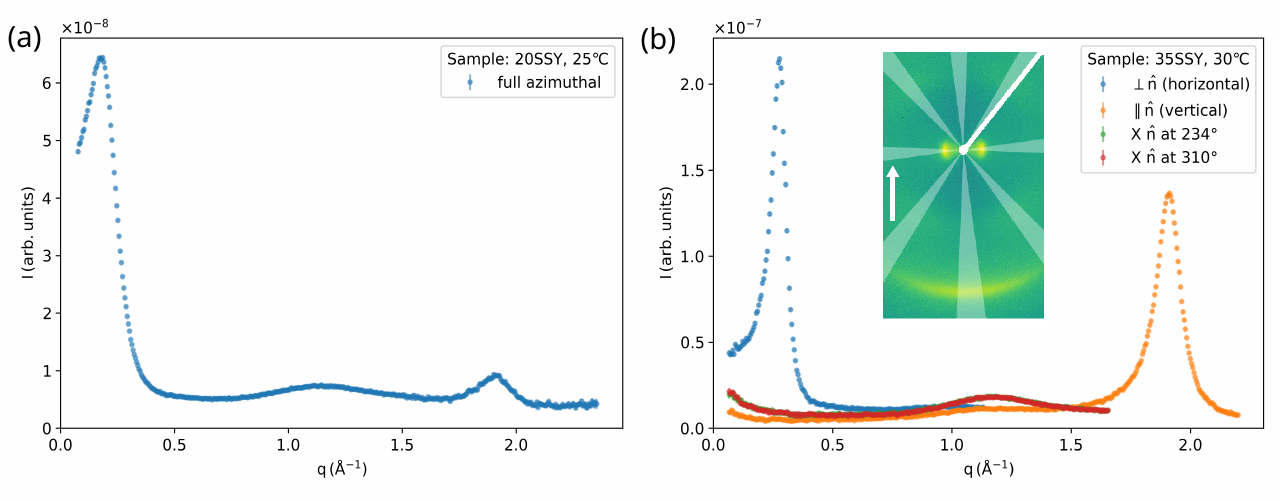}
	\caption{X-ray scattering curves from the images of Figure \ref{fig:waxs:2D}. (a) Isotropic phase of sample 20SSY at $T = 25.0\,^{\circ} C;$ (b) oriented nematic phase of sample 35SSY at $T = 30.0\,^{\circ} C.$ Inset: For the oriented N phase, the sectors used for the azimuthal averages are highlighted in white over the WAXS image. The white arrow represent the N phase director, $\hat{n}.$}
	\label{fig:waxs:integracoes}
\end{figure}

\begin{figure}[ht]
	\centering
	\def\svgwidth{0.5\linewidth}
	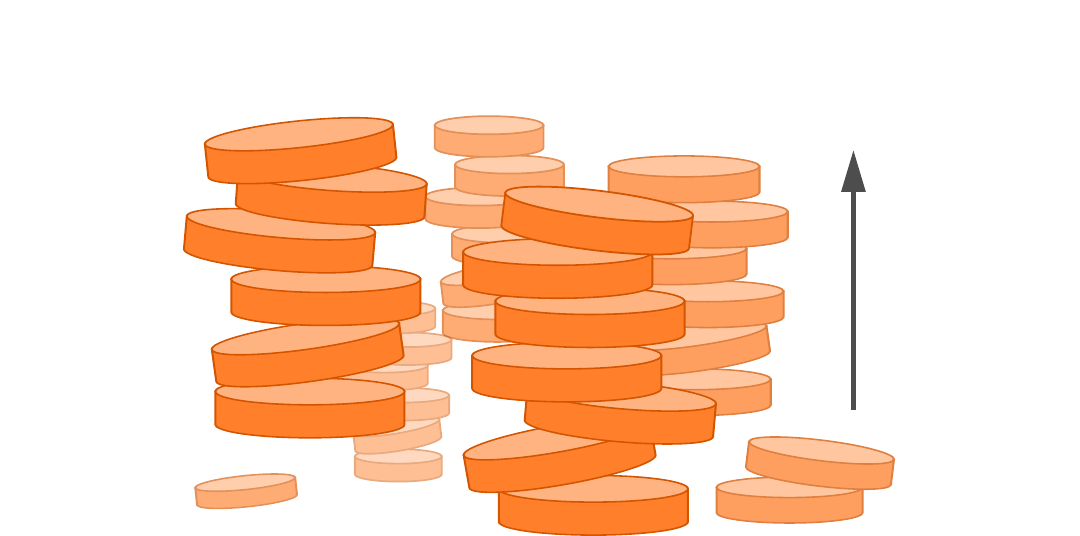
	\caption{Sketch of the molecular stacks organization in the lyotropic chromonic nematic phase. The structural parameters obtained from WAXS are identified. $d_{\parallel}$ is the average intermolecular distance in a stack, and $d_{\perp}$ is the interstacks average distance, i.e., the average distance between adjacent stacks. The $\hat{n}$ defines the phase director.}
	\label{fig:cromonico:esquema}
\end{figure}

From them, the scattered intensity as a function of $q$ curves are obtained, Figure \ref{fig:waxs:integracoes}. Three bands are identified in Figures \ref{fig:waxs:2D} and \ref{fig:waxs:integracoes}, at $q_{\perp} \approx 0.18\,\si{\angstrom}^{-1},$ $q_X \approx 1.1\,\si{\angstrom}^{-1},$ and $q_{\parallel} \approx 1.9\,\si{\angstrom}^{-1}.$ In particular, the bands at $q_X$ in the oriented N phases are at about $ \pm 37 ^{\circ}$ from the director orientation direction.
The $\parallel$ band is related to the distance between SSY molecules in a stack, $d_{\parallel}.$ The band at $q_{\perp}$ is related to the average distance between molecular aggregates, $d_{\perp}.$ Figure \ref{fig:cromonico:esquema} sketches a nematic structure just to show the labels of $d_{\perp}$ and $d_\parallel.$

\begin{figure}[H]
\centering
\includegraphics[scale=0.5]{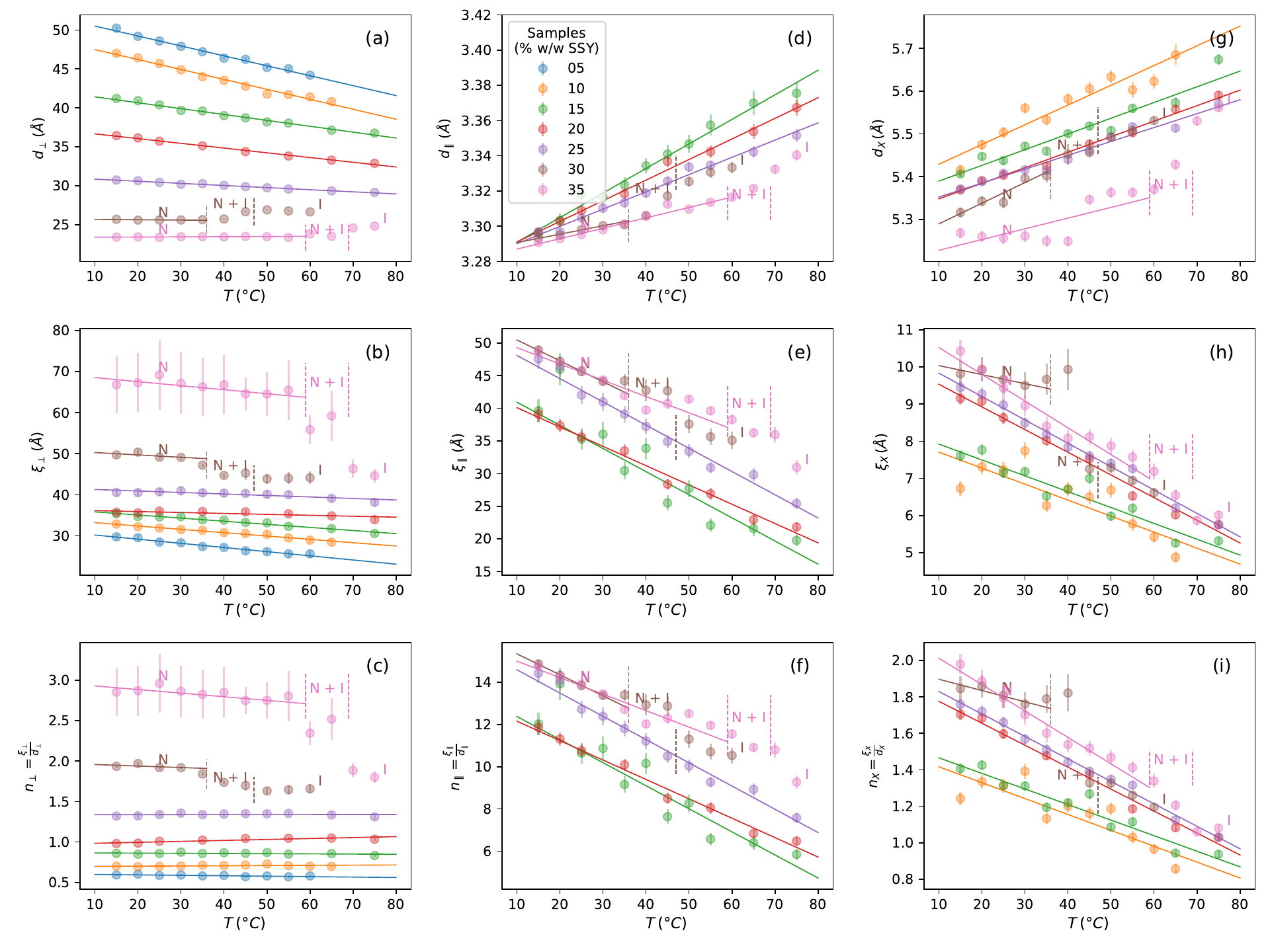}
\caption{Structural parameters as a function of temperature, measured from WAXS images for samples of different mass fractions, $w_{SSY}.$ (a) $d_{\perp}$ is the average interstack distance; (d) $d_{\parallel}$ is the average intermolecular distance in a stack; (g) $d_X$ is the average distance corresponding to the diagonal peaks; (b) $\xi_{\perp}$ is the interstack correlation length; (e) $\xi_{\parallel}$ is the intrastack correlation length along the stacking direction; (h) $\xi_X;$ (c) $n_{\perp} = \frac{\xi_{\perp}}{d_{\perp}};$ (f) $n_{\parallel} = \frac{\xi_{\parallel}}{d_{\parallel}};$ and (i) $n_X = \frac{\xi_X}{d_X}.$ The colored solid lines are linear fits to the data points of corresponding color. The dashed vertical lines shows the temperatures of the phase transitions from the nematic phase (N) to the coexistence region (N+I) and from the coexistence to the isotropic phase (I).}
\label{fig:drx:temperatura}
\end{figure}

\begin{figure}[H]
\centering
\includegraphics[scale=0.5]{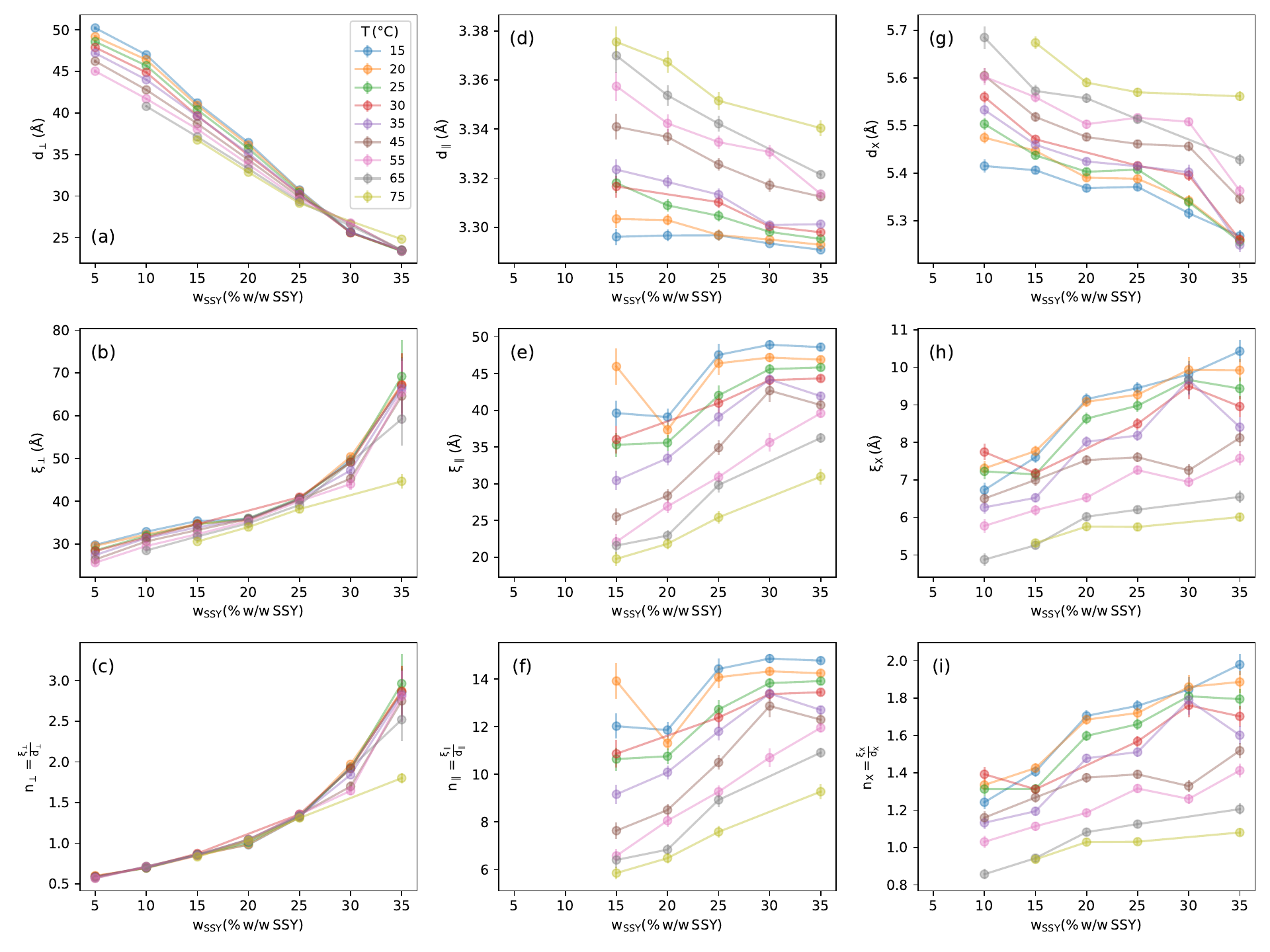}
\caption{Structural parameters as a function of samples mass fractions, $w_{SSY},$ measured from WAXS images for different temperatures. (a) $d_{\perp}$ is the average interstack distance; (b) $d_X$ is the average distance corresponding to the diagonal peaks; (c) $d_{\parallel}$ is the average intermolecular distance in a stack; (d) $\xi_{\perp}$ is the interstack correlation length; (e) $\xi_X$ ; (f) $\xi_{\parallel}$ is the intrastack correlation length along the stacking direction; (g) $n_{\perp} = \frac{\xi_{\perp}}{d_{\perp}}$ ;(h) $n_X = \frac{\xi_X}{d_X}$ ; and (i) $n_{\parallel} = \frac{\xi_{\parallel}}{d_{\parallel}}$. The colored lines are linear fits to the data points of corresponding color.}
\label{fig:drx:concentracao}
\end{figure}

Let us first analyze the isotropic phase as a function of the temperature and $w_{SSY}.$ $d_{\perp}$ represents the mean distance between the centers of mass of the molecular aggregates. Increasing the $w_{SSY},$ $d_{\perp}$ decreases, Figures \ref{fig:drx:temperatura} and \ref{fig:drx:concentracao} (a). From the full-width at half maximum (FWHM) of the $\perp$ peaks, the coherence lengths $\xi_{\perp}$ were evaluated and the average number of aggregates that originates this band $ n_\perp =  \frac{\xi_\perp}{d_\perp},$ shown in Figures \ref{fig:drx:temperatura} and \ref{fig:drx:concentracao} (b) and (c), respectively. In the concentration range investigated, $n_\perp$ increases  $(0.5 < n_\perp < 1.5),$ and is temperature independent. An interesting result is observed in the temperature dependence of $d_\perp,$ for fixed $w_{SSY}:$ $d_\perp$ decreases as a function of the temperature. We will be back to this point in the following. 

Along the direction parallel to the cylindrical aggregates, $d_\parallel$ informs about the distance between neighbor molecules, Figures \ref{fig:drx:concentracao} (d), (e), and (f). As a function of temperature, at fixed $w_{SSY},$ the number of molecules per aggregates decreases, Figure \ref{fig:drx:concentracao}(f). This result is consistent with the decrease of $d_\perp$ as the temperature increases, Figure \ref{fig:drx:concentracao}(a), as additional free molecules and small aggregates are present in the bulk of the mesophase.

The band at $q_X$ seems to be related to the electronic density contrast in the borders of the molecules in the stacks. The characteristic distance associated to $q_X$ is $\approx 55\, \si{\angstrom}$ and increases as a function of the temperature, Figures \ref{fig:drx:concentracao}(g), (h), and (i). The number of scatterers that originates this broad band is about 1.5 and decrease as a function of increasing temperature. This result indicate that the molecular ordering in the stacks lowers.

In the nematic phase, the average distance between the aggregates $d_\perp$ is about $25 \, \si{\angstrom},$ Figure \ref{fig:drx:concentracao}(a). From the correlation distance $\xi_{\perp} ,$ the average number of stacks in the correlation volume is about $2,$ Figure \ref{fig:drx:concentracao}(c). The analysis of the band at $q_{\parallel}$ reveals that the stacks have about $14$ molecules, and this number decreases with the increase of the temperature, Figures \ref{fig:drx:concentracao}(d)-(f).

\subsection{Nonlinear Index of Refraction}

\begin{figure}[H]
\centering
\includegraphics[scale=0.4]{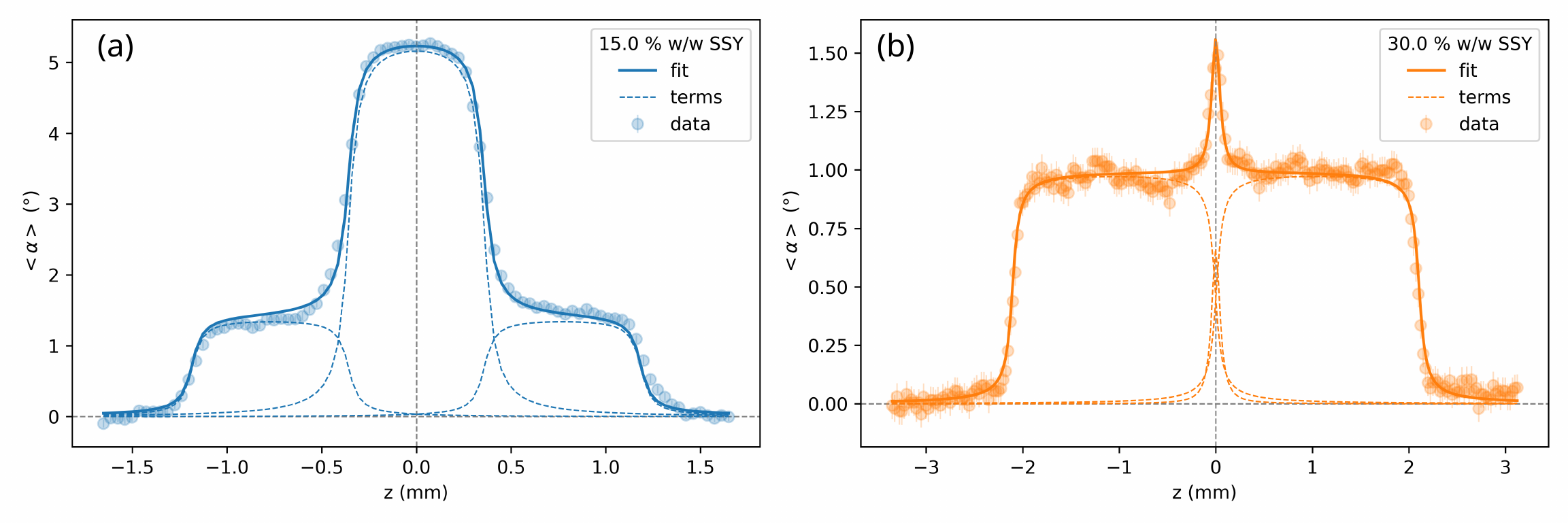}
\caption{Polarization ellipse rotation angle $<\alpha>,$ in degrees, as a function of sample position $z,$ in mm. $z=0$ is the focus position. The solid lines are fits of equation \ref{eq:ner:sample}. The samples were placed in (a) quartz cuvette or in (b) quartz sample holders. The dotted lines show the terms corresponding to each holder window and to the sample. See equation \ref{eq:ner:sample}. Samples mass fraction $w_{SSY},$ in $\%\,\text{w/w}$ SSY for different pulse width $\tau:$  (a) $w_{SSY} = 15.0\,\%, \tau = 550\,fs$ and (b) $w_{SSY} = 30.0\,\%, \tau = 67\,fs.$}
\label{fig:ner:curves}
\end{figure}

\begin{figure}[H]
\centering
\includegraphics[scale=0.4]{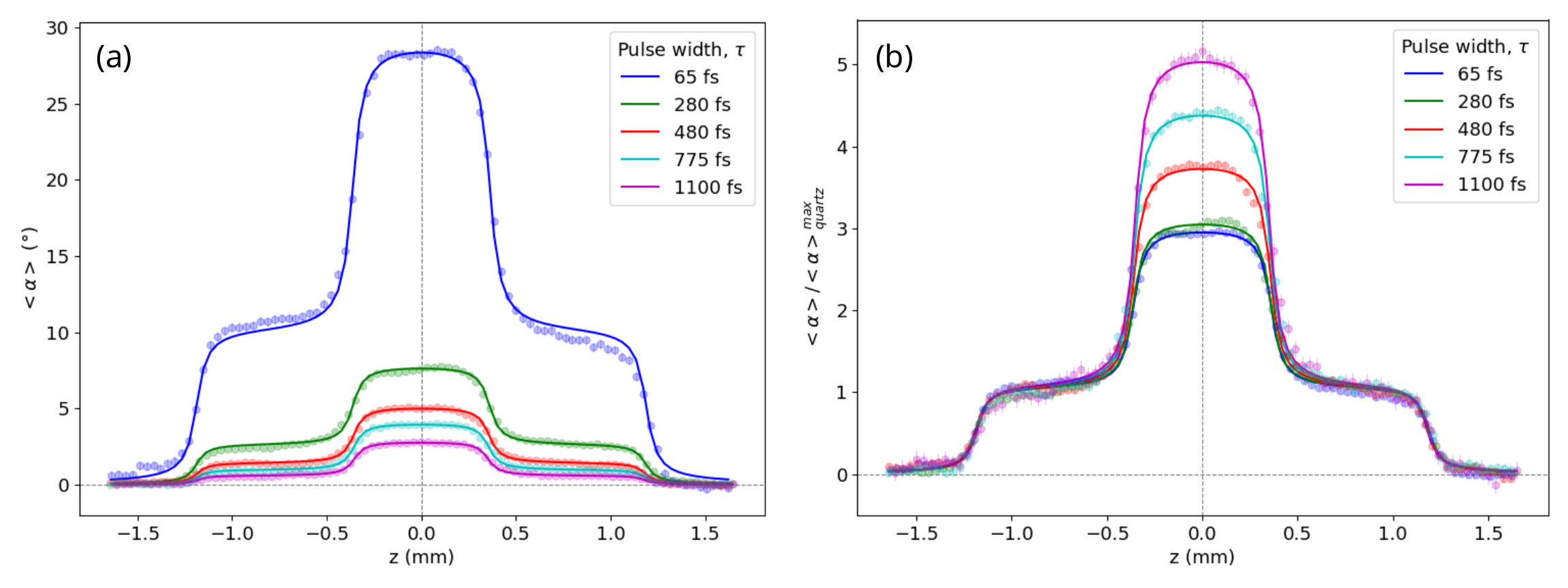}
\caption{Polarization ellipse rotation angle $<\alpha>$ as a function of sample position $z$ for sample 15SSY for different pulse widths. (a) NER signals measured with the same average laser power. (b) The same signals were normalized by the amplitude of the quartz signal from the sample holder. The solid lines are fits of Equation \ref{eq:ner:sample} to the data points of the corresponding color.}
\label{fig:ner:pulselength}
\end{figure}

As discussed elsewhere \cite{miguez_measurement_2017, fernandes_small_2021}, the NER experiment allows the determination of the index of refraction associated to two physical processes: a fast one associated to the electronic transitions in the molecules, and another, named slow, associated to molecular and/or stacks reorientation. The measurements consist in obtain the rotation angle $<\alpha>$ for different laser pulse widths as a function of the position $z$ of the sample along the beam propagation direction, Figure \ref{fig:ner:curves}. The NER signal from the sample is the highest peak around $z=0,$ and those in both sides of the ``main peak'' are due to the sample holder quartz windows. These holder signals refer to $n_0$ and $n_2$ of the quartz and are used as reference to obtain $I_0$ and $z_0$ from the NER setup. With short pulses $(\tau << \tau_0),$  the $n_{2,fast}$ was calculated, and with $\tau > \tau_0,$ one gets $n_{2,slow},$ where $\tau_0$ is a characteristic response time of the sample non-electronic or ``slow'' nonlinearities. Figure \ref{fig:ner:curves} shows these typical results, where $z=0$ is the focus position. Equations \ref{eq:ner:angle} and \ref{eq:ner:sample} are used to fit the NER experimental results. Figure \ref{fig:ner:pulselength} shows typical NER results as a function of pulse width, $\tau.$ 

Now, we use an empirical expression for the $B_{eff}$ coefficient as a function of pulse width $\tau$ \cite{miguez_measurement_2017}:

\begin{equation}
B_{eff} = B_{fast} + B_{slow} \left[ 1-\exp \left( \frac{-\tau}{\tau_0} \right)\right], 
\label{eq:ner:model}
\end{equation}
where $\tau_0$ is a characteristic response time of the sample non-electronic nonlinearities, $B_{fast}$ is the $B$ coefficient due to molecular electronic processes, and $B_{slow}$ is due to molecular reorientational processes.

Relations of $B_{fast}$ with $n_{2,fast}$ and of $B_{slow}$ with $n_{2,slow}$ are, respectively \cite{miguez_measurement_2017}:

\begin{equation}
n_{2,fast} = \frac{3}{8}\frac{B_{fast}}{n_0^2 \epsilon_0 c},\,\,
n_{2,slow} = \frac{1}{6}\frac{B_{slow}}{n_0^2 \epsilon_0 c}.
\label{eq:ner:n2s}
\end{equation}

\begin{figure}[H]
\centering
\includegraphics[scale=1]{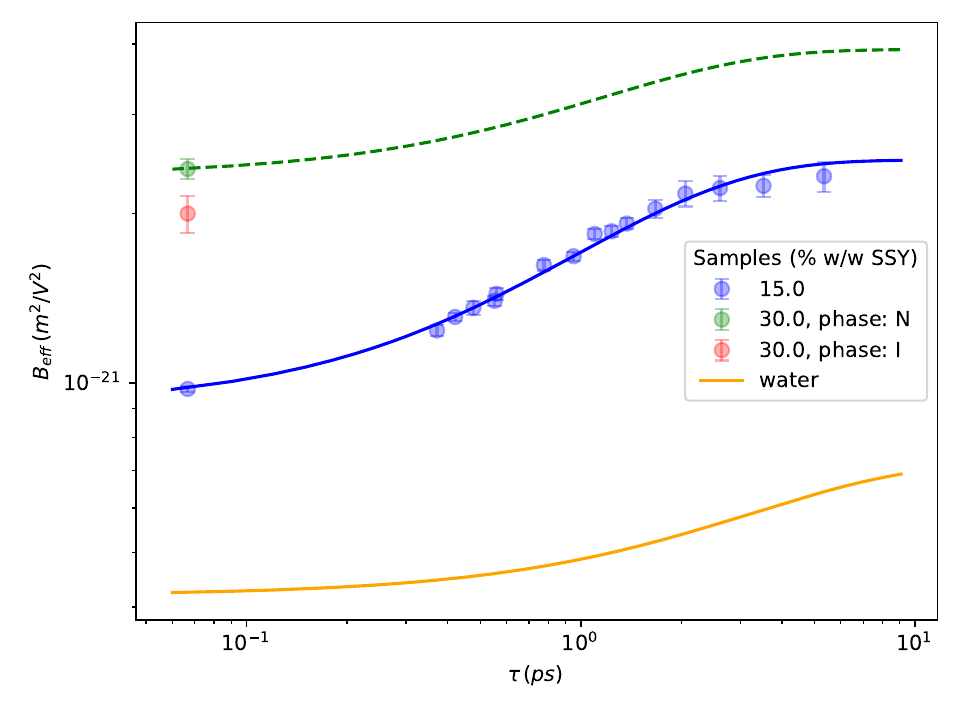}
\caption{Experimental $B_{eff}$ values of chromonic samples as a function of pulse width, $\tau.$ The $B_{eff}$ data points were obtained from fitting NER curves with Equation \ref{eq:ner:sample}. The solid lines are fits of the empirical expression given by Equation \ref{eq:ner:model}. The water curve were calculated from the data given in Table \ref{tab:ner}, taken from \cite{miguez_measurement_2017}. Except for sample 30SSY in isotropic phase, which was measured at $50\,^{\circ}C,$ all other measurements were carried out at $25\,^{\circ}C.$ The dashed line represents an evolution of the $B_{eff}$ dependence on $\tau,$ assuming the same behavior observed for the isotropic phase of sample 15SSY.}
\label{fig:ner:b_eff}
\end{figure}

Figure \ref{fig:ner:b_eff} shows the parameter $B_{eff}$ as a function of the pulse width, $\tau.$ $B_{eff}$ values were obtained from the fit of the NER experimental curves with Equation \ref{eq:ner:sample}, as shown in Figure \ref{fig:ner:pulselength} for different pulse widths. Assuming the empirical expression of Equation \ref{eq:ner:model}, $B_{fast}$ and $B_{slow}$ are obtained from the fitting to the experimental data. The corresponding nonlinear indices of refraction $n_{2,fast}$ and $n_{2,slow}$ were calculated from Equations \ref{eq:ner:n2s}, and the results are shown in Table \ref{tab:ner}. 

\begin{figure}[H]
\centering
\includegraphics[scale=0.7]{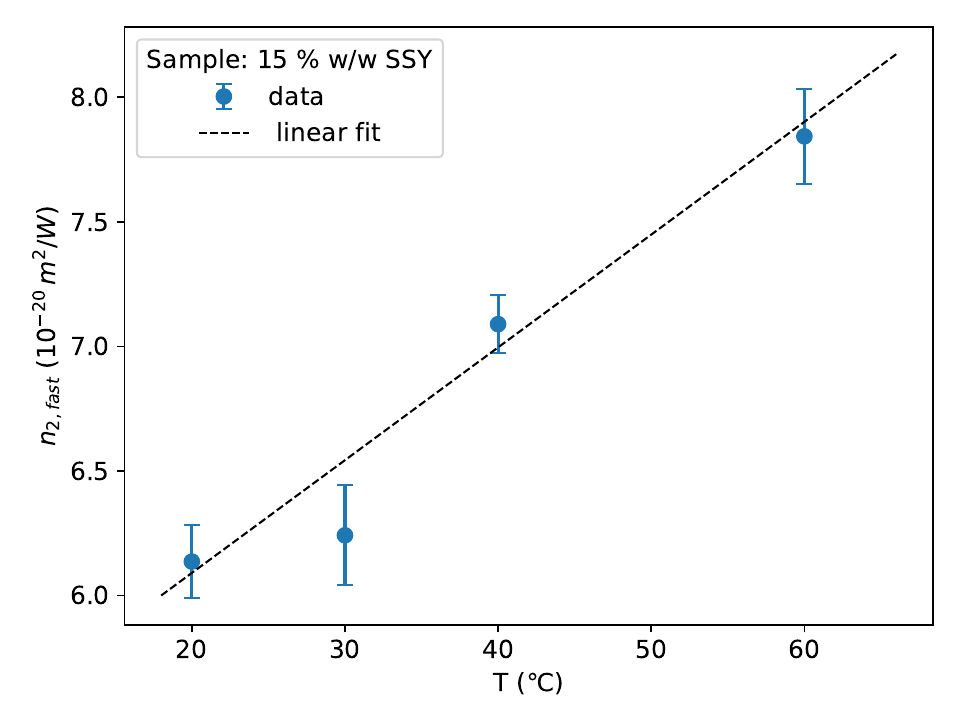}
\caption{Experimental $n_{2,fast}$ as a function of temperature, $T,$ for sample 15SSY, measured with pulse width $\tau = 67 \,fs.$ Dashed line is a linear fit used just as a guide.}
\label{fig:ner:Beff_T}
\end{figure}

In the isotropic phase of sample 15SSY at $25\, ^{\circ}C ,$ $n_{2,fast} \approx 6.5 \times 10^{-20} m^2/W$ and $n_{2,slow} \approx 5 \times 10^{-20} m^2/W.$ The temperature variation of $n_{2,fast}$ is shown in Figure \ref{fig:ner:Beff_T}. In the investigated temperature range of $40\,^{\circ}C,$ $n_{2,fast}$ increased about $28\%.$

\begin{table}[H]
\caption{\label{tab:ner} Parameters of the third-order nonlinear responses from the fitting of Equation \ref{eq:ner:model} to the experimental data. The data for water were taken from \cite{miguez_measurement_2017}. $\tau_0$ is in $ps,$ $B$ is in $10^{-21}\,m^2/V^2,$ and $n_2$ is in $10^{-20}\,m^2/W.$ The samples are identified by their percentage by mass of SSY, $w_{SSY}$. ``30SSY, phase I'' was measured at $50\,^{\circ}C,$ all the others at $25\,^{\circ}C.$}
\begin{ruledtabular}
\begin{tabular}{cccccc}
Sample & $\tau_0$ & $B_{fast}$ & $B_{slow}$ & $n_{2,fast}$ & $n_{2,slow}$ \\ \hline
water & $4.0$ & $0.42$ & $0.3$ & $3.35$ & $1.06$ \\
15SSY & $1.29 \pm 0.10$ & $0.893 \pm 0.011$ & $1.55 \pm 0.07$ & $6.47 \pm 0.08$ & $4.99 \pm 0.22$  \\
30SSY, phase I & -  & $2.00 \pm 0.15$ & - & $14.5 \pm 1.1$ & - \\
30SSY, phase N & $1.3\,^*$  & $2.40 \pm 0.10$ & $1.55\,^*$ & $17.4 \pm 0.7$ & $5\,^*$ \\       
\end{tabular}
\end{ruledtabular}
{\raggedright $^*$Assuming the same behavior observed for the 15SSY sample. \par}
\end{table}

In the case of the nematic sample, despite the planar alignment of the director, we did not observed anisotropy of $n_{2,fast}.$ In this experiment, the polarization of the laser beam was oriented parallel and perpendicular to the director, but no reliable differences were obtained in the NER curves. Therefore, the $n_{2,fast} = 17.4 \times 10^{-20}\,m^2/W$ is a mean value of this parameter for the 30SSY nematic mesophase as $25\,^{\circ}C.$ 

Increasing the temperature of the 30SSY to $50\,^{\circ}C,$ the isotropic phase was reached, and the measured $n_{2,fast} = 14.5 \times 10^{-20}\,m^2/W.$ Figure \ref{fig:drx:concentracao}(f) shows that the average number of molecules per stack in the 30SSY at $25\,^{\circ}C$ is about $14,$ and around $50\,^{\circ}C,$ this number decreases to about $10.$ In percentage terms, $n_{2,fast}$ decreased about $17\,\%,$ but remains higher than the typical values obtained in the isotropic phase in the mixture 15SSY. 

Interestingly, normalizing the $n_{2,fast}$ by the mass percentage of SSY in both isotropic phases, at the same temperature of $50\,^{\circ}C,$ the difference of $n_{2,fast}$ is smaller than $2\,\%.$ The average number of molecules in the stack at $50\,^{\circ}C$ for 15SSY is about $8,$ near the number in 30SSY for the same temperature. This reinforces the molecular characteristic of $n_{2,fast}$ measured in the distinct isotropic phases 15SSY and 30SSY. 

To evaluate the $n_{2,slow}$ in the nematic phase of 30SSY at $25\,^{\circ}C,$ it was assumed the same behavior of $B_{eff}$ as a function of the pulse width $\tau$ as that observed for the isotropic phase of 15SSY at $25\,^{\circ}C.$ The dashed line shown in Figure \ref{fig:ner:b_eff} was calculated with this assumption, starting with the $B_{eff}$ measured with 30SSY at $\tau = 67\,fs.$ With this procedure, we evaluated $n_{2,slow} \approx 5 \times 10^{-20} m^2/W,$ which is about the same value of $n_{2,slow}$ of the isotropic phase measured in the 15SSY sample at $25\,^{\circ}C.$ This parameter is associated with molecular and stacks present in the mixture. From the structural results of Figures \ref{fig:drx:temperatura} and \ref{fig:drx:concentracao}, the average number of molecules in the stack in 15SSY is about $11$ and in 30SSY is about $14,$ for both at $25\,^{\circ}C.$ The same value of $n_{2,slow}$ obtained in both mixtures, despite the fact that they are different mesophases, N and I, seems to indicate that the nonlinear effect related to the stacks is, mainly, influenced by the number of molecules in the stack than in the particular mesophase (obviously, in the SSY relative concentrations investigated).

\section{ Conclusions }

The nanostructures of the prototypical chromonic system made of aqueous solutions of sunset yellow (SSY) azo dye were characterized by wide angle X-ray scattering. We investigated samples across the concentration range between $5$ and $35\,\%\,\text{w/w}$ SSY and the temperature range between $15$ and $75\,^{\circ}C.$

The nonlinear refractive index $n_2$ of SSY samples in the isotropic and nematic phases were measured and discussed based on the structural characteristics of the systems. The nonlinear polarization ellipse rotation technique was shown to be a powerful and reliable technique to measure the fast and slow nonlinear refractive indices of the mesophases associated to electronic and orientational processes. For the isotropic phase of sample $15\,\%\,\text{w/w}$ SSY, $n_2$ associated to fast molecular electronic processes $( \lesssim 100\,fs),$ $n_{2,fast},$ and slow molecular reorientational processes $( \gtrsim 1\, ps),$ $n_{2,slow},$ both at $25\,^{\circ}C,$ are $n_{2,fast} = (6.47 \pm 0.08) \times 10^{-20}\,m^2/W,$ and $n_{2,slow} = (4.99 \pm 0.22) \times 10^{-20}\,m^2/W,$ respectively. Also, the change of $n_{2,fast}$ with temperature is approximately linear and increased $28\%$ between $20$ and $60\,^{\circ}C.$ For the $30\,\%\,\text{w/w}$ SSY sample, the $n_{2,fast}$ of the aligned nematic and isotropic phases are $n_{2,fast} = (17.4 \pm 0.7) \times 10^{-20}\,m^2/W,$ at $25\,^{\circ}C,$ and $n_{2,fast} = (14.5 \pm 1.1) \times 10^{-20}\,m^2/W,$ at $50\,^{\circ}C,$ respectively. We found that $n_{2,fast}$ doubled in value from the isotropic phases of the $15$ to the $30\,\%\,\text{w/w}$ samples, proportionally to the increase in mass fraction. On the other hand, $n_{2,fast}$ for the nematic phase of $30\,\%\,\text{w/w}$ sample is higher than the double of the corresponding value for the $15\,\%\,\text{w/w}$ sample, showing an effect in $n_2$ beyond that associated to the increase in concentration of SSY molecules, and, probably, associated to the orientational order of this phase. No anisotropies were observed in $n_{2,fast}$ for the nematic phase using our experimental technique.

The values of $n_{2,slow}$ measured in the isotropic mesophase (15 \%\,\text{w/w} SSY) and that evaluated in the nematic mesophase (30 \%\,\text{w/w} SSY) were shown to be about the same. These results seems to indicate that the nonlinear effect associated to the reorientation of the stacks depends more on the structure of them, than the mesophase investigated, at least in the concentrations investigated.

% If you have acknowledgments, this puts in the proper section head.
\begin{acknowledgments}
This study was financed, in part, by the São Paulo Research Foundation (FAPESP), Brasil. Process Number 2023/10843-7.

\end{acknowledgments}

% Create the reference section using BibTeX:
\bibliography{ bibliography }

\end{document}

%% file: Figures/ner_setup.pdf_tex
%% Creator: Inkscape 1.1.2 (0a00cf5339, 2022-02-04), www.inkscape.org
%% PDF/EPS/PS + LaTeX output extension by Johan Engelen, 2010
%% Accompanies image file 'ner_setup.pdf' (pdf, eps, ps)
%%
%% To include the image in your LaTeX document, write
%%   \input{<filename>.pdf_tex}
%%  instead of
%%   \includegraphics{<filename>.pdf}
%% To scale the image, write
%%   \def\svgwidth{<desired width>}
%%   \input{<filename>.pdf_tex}
%%  instead of
%%   \includegraphics[width=<desired width>]{<filename>.pdf}
%%
%% Images with a different path to the parent latex file can
%% be accessed with the `import' package (which may need to be
%% installed) using
%%   \usepackage{import}
%% in the preamble, and then including the image with
%%   \import{<path to file>}{<filename>.pdf_tex}
%% Alternatively, one can specify
%%   \graphicspath{{<path to file>/}}
%% 
%% For more information, please see info/svg-inkscape on CTAN:
%%   http://tug.ctan.org/tex-archive/info/svg-inkscape
%%
\begingroup%
  \makeatletter%
  \providecommand\color[2][]{%
    \errmessage{(Inkscape) Color is used for the text in Inkscape, but the package 'color.sty' is not loaded}%
    \renewcommand\color[2][]{}%
  }%
  \providecommand\transparent[1]{%
    \errmessage{(Inkscape) Transparency is used (non-zero) for the text in Inkscape, but the package 'transparent.sty' is not loaded}%
    \renewcommand\transparent[1]{}%
  }%
  \providecommand\rotatebox[2]{#2}%
  \newcommand*\fsize{\dimexpr\f@size pt\relax}%
  \newcommand*\lineheight[1]{\fontsize{\fsize}{#1\fsize}\selectfont}%
  \ifx\svgwidth\undefined%
    \setlength{\unitlength}{498.42916173bp}%
    \ifx\svgscale\undefined%
      \relax%
    \else%
      \setlength{\unitlength}{\unitlength * \real{\svgscale}}%
    \fi%
  \else%
    \setlength{\unitlength}{\svgwidth}%
  \fi%
  \global\let\svgwidth\undefined%
  \global\let\svgscale\undefined%
  \makeatother%
  \begin{picture}(1,0.57304618)%
    \lineheight{1}%
    \setlength\tabcolsep{0pt}%
    \put(0,0){\includegraphics[width=\unitlength,page=1]{ner_setup.pdf}}%
    \put(0.48810971,0.29168198){\color[rgb]{0.2,0.2,0.2}\makebox(0,0)[lt]{\smash{\begin{tabular}[t]{l}0\\\end{tabular}}}}%
    \put(0.55711124,0.34048054){\color[rgb]{0.2,0.2,0.2}\makebox(0,0)[lt]{\smash{\begin{tabular}[t]{l}z\\\end{tabular}}}}%
    \put(0.81610796,0.44405418){\color[rgb]{0,0,0}\makebox(0,0)[lt]{\smash{\begin{tabular}[t]{l}photodetector\end{tabular}}}}%
    \put(0.45907411,0.45958452){\color[rgb]{0,0,0}\makebox(0,0)[lt]{\lineheight{0.80000001}\smash{\begin{tabular}[t]{l}sample in\\cuvette \end{tabular}}}}%
    \put(0.08328982,0.40076579){\color[rgb]{0,0,0}\makebox(0,0)[lt]{\lineheight{0.80000001}\smash{\begin{tabular}[t]{l}fs laser\\$\lambda = 800\,nm$\end{tabular}}}}%
    \put(0,0){\includegraphics[width=\unitlength,page=2]{ner_setup.pdf}}%
    \put(0.29293629,0.33594826){\color[rgb]{0,0,0}\makebox(0,0)[lt]{\smash{\begin{tabular}[t]{l}objective\\lens \end{tabular}}}}%
    \put(0,0){\includegraphics[width=\unitlength,page=3]{ner_setup.pdf}}%
    \put(0.65350783,0.33623289){\color[rgb]{0,0,0}\makebox(0,0)[lt]{\smash{\begin{tabular}[t]{l}objective\\lens \end{tabular}}}}%
    \put(0.42272247,0.26770236){\color[rgb]{0,0,0}\makebox(0,0)[lt]{\smash{\begin{tabular}[t]{l}translation stage\end{tabular}}}}%
    \put(0.71187301,0.49692939){\color[rgb]{0,0,0}\makebox(0,0)[lt]{\lineheight{0.80000001}\smash{\begin{tabular}[t]{l}rotating\\polarizer\\(40 Hz)\end{tabular}}}}%
    \put(0.21972948,0.46170299){\color[rgb]{0,0,0}\makebox(0,0)[lt]{\lineheight{0.89999998}\smash{\begin{tabular}[t]{l}$\lambda/4$ waveplate\end{tabular}}}}%
    \put(0,0){\includegraphics[width=\unitlength,page=4]{ner_setup.pdf}}%
    \put(0.73004561,0.26190548){\color[rgb]{0,0,0}\makebox(0,0)[lt]{\lineheight{0.69999999}\smash{\begin{tabular}[t]{l}dual phase\\lock-in amplifier\end{tabular}}}}%
    \put(0.75266458,0.10219714){\color[rgb]{0,0,0}\makebox(0,0)[lt]{\smash{\begin{tabular}[t]{l}oscilloscope\end{tabular}}}}%
    \put(0.75540769,0.29088468){\color[rgb]{0,0,0}\makebox(0,0)[lt]{\smash{\begin{tabular}[t]{l}ref\end{tabular}}}}%
    \put(0.81415011,0.29078741){\color[rgb]{0,0,0}\makebox(0,0)[lt]{\smash{\begin{tabular}[t]{l}input\end{tabular}}}}%
    \put(0.28152586,0.13639863){\color[rgb]{0.2,0.2,0.2}\makebox(0,0)[lt]{\smash{\begin{tabular}[t]{l}x\\\end{tabular}}}}%
    \put(0.14278081,0.27608772){\color[rgb]{0.2,0.2,0.2}\makebox(0,0)[lt]{\smash{\begin{tabular}[t]{l}y\\\end{tabular}}}}%
    \put(0,0){\includegraphics[width=\unitlength,page=5]{ner_setup.pdf}}%
    \put(0.25727966,0.23688354){\color[rgb]{0,0,0}\makebox(0,0)[lt]{\smash{\begin{tabular}[t]{l}$\alpha$\end{tabular}}}}%
    \put(0.18302374,0.05928235){\color[rgb]{0,0,0}\makebox(0,0)[lt]{\smash{\begin{tabular}[t]{l}polarization ellipse\end{tabular}}}}%
    \put(0.08796552,0.21318272){\color[rgb]{1,0.4,0}\makebox(0,0)[lt]{\smash{\begin{tabular}[t]{l}incident\end{tabular}}}}%
    \put(0.19469885,0.11747714){\color[rgb]{0,0,1}\makebox(0,0)[lt]{\smash{\begin{tabular}[t]{l}transmitted\end{tabular}}}}%
  \end{picture}%
\endgroup%

%% file: Figures/esquema_cromonicos.pdf_tex
%% Creator: Inkscape 1.1.2 (0a00cf5339, 2022-02-04), www.inkscape.org
%% PDF/EPS/PS + LaTeX output extension by Johan Engelen, 2010
%% Accompanies image file 'esquema_cromonicos.pdf' (pdf, eps, ps)
%%
%% To include the image in your LaTeX document, write
%%   \input{<filename>.pdf_tex}
%%  instead of
%%   \includegraphics{<filename>.pdf}
%% To scale the image, write
%%   \def\svgwidth{<desired width>}
%%   \input{<filename>.pdf_tex}
%%  instead of
%%   \includegraphics[width=<desired width>]{<filename>.pdf}
%%
%% Images with a different path to the parent latex file can
%% be accessed with the `import' package (which may need to be
%% installed) using
%%   \usepackage{import}
%% in the preamble, and then including the image with
%%   \import{<path to file>}{<filename>.pdf_tex}
%% Alternatively, one can specify
%%   \graphicspath{{<path to file>/}}
%% 
%% For more information, please see info/svg-inkscape on CTAN:
%%   http://tug.ctan.org/tex-archive/info/svg-inkscape
%%
\begingroup%
  \makeatletter%
  \providecommand\color[2][]{%
    \errmessage{(Inkscape) Color is used for the text in Inkscape, but the package 'color.sty' is not loaded}%
    \renewcommand\color[2][]{}%
  }%
  \providecommand\transparent[1]{%
    \errmessage{(Inkscape) Transparency is used (non-zero) for the text in Inkscape, but the package 'transparent.sty' is not loaded}%
    \renewcommand\transparent[1]{}%
  }%
  \providecommand\rotatebox[2]{#2}%
  \newcommand*\fsize{\dimexpr\f@size pt\relax}%
  \newcommand*\lineheight[1]{\fontsize{\fsize}{#1\fsize}\selectfont}%
  \ifx\svgwidth\undefined%
    \setlength{\unitlength}{518.1221385bp}%
    \ifx\svgscale\undefined%
      \relax%
    \else%
      \setlength{\unitlength}{\unitlength * \real{\svgscale}}%
    \fi%
  \else%
    \setlength{\unitlength}{\svgwidth}%
  \fi%
  \global\let\svgwidth\undefined%
  \global\let\svgscale\undefined%
  \makeatother%
  \begin{picture}(1,0.49645197)%
    \lineheight{1}%
    \setlength\tabcolsep{0pt}%
    \put(0,0){\includegraphics[width=\unitlength,page=1]{esquema_cromonicos.pdf}}%
    \put(0.81410625,0.32890963){\color[rgb]{0.2,0.2,0.2}\makebox(0,0)[lt]{\lineheight{1.25}\smash{\begin{tabular}[t]{l}$\hat{n}$\end{tabular}}}}%
    \put(0,0){\includegraphics[width=\unitlength,page=2]{esquema_cromonicos.pdf}}%
    \put(0.37707893,0.46418162){\color[rgb]{0.2,0.2,0.2}\makebox(0,0)[lt]{\lineheight{1.25}\smash{\begin{tabular}[t]{l}$d_{\perp}$\end{tabular}}}}%
    \put(-0.00263258,0.32330942){\color[rgb]{0.30196078,0.30196078,0.30196078}\makebox(0,0)[lt]{\lineheight{1.25}\smash{\begin{tabular}[t]{l}$d_{\parallel}$\end{tabular}}}}%
    \put(0,0){\includegraphics[width=\unitlength,page=3]{esquema_cromonicos.pdf}}%
  \end{picture}%
\endgroup%